# What Types of Chemical Problems Benefit from Density-Corrected DFT? A Probe Using an Extensive and Chemically Diverse Test Suite

Golokesh Santra and Jan M.L. Martin*

**ABSTRACT:** For the large and chemically diverse GMTKN55 benchmark suite, we have studied the performance of density-corrected density functional theory (HF-DFT), compared to self-consistent DFT, for several pure and hybrid GGA and meta-GGA exchange–correlation (XC) functionals (PBE, BLYP, TPSS, and SCAN) as a function of the percentage of HF exchange in the hybrid. The D4 empirical dispersion correction has been added throughout. For subsets dominated by dynamical correlation, HF-DFT is highly beneficial, particularly at low HF exchange percentages. This is especially true for noncovalent interactions where the electrostatic component is dominant, such as hydrogen and halogen bonds: for $\pi$-stacking, HF-DFT is detrimental. For subsets with significant nondynamical correlation (i.e., where a Hartree–Fock determinant is not a good zero-order wavefunction), HF-DFT may do more harm than good. While the self-consistent series show optima at or near 37.5% (i.e., 3/8) for all four XC functionals—consistent with Grimme's proposal of the PBE38 functional—HF-B$n$LYP-D4, HF-PBE$n$-D4, and HF-TPSS$n$-D4 all exhibit minima nearer 25% (i.e., 1/4) as the use of HF orbitals greatly mitigates the error at 25% for barrier heights. Intriguingly, for HF-SCAN$n$-D4, the minimum is near 10%, but the weighted mean absolute error (WTMAD2) for GMTKN55 is only barely lower than that for HF-SCAN-D4 (i.e., where the post-HF step is a pure meta-GGA). The latter becomes an attractive option, only slightly more costly than pure Hartree–Fock, and devoid of adjustable parameters other than the three in the dispersion correction. Moreover, its WTMAD2 is only surpassed by the highly empirical M06-2X and by the combinatorially optimized empirical range-separated hybrids $\omega$B97X-V and $\omega$B97M-V.

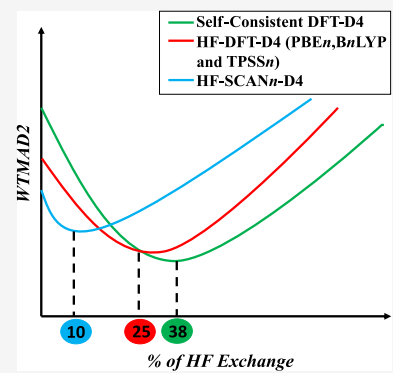

## INTRODUCTION

In a review with the provocative title "The importance of being inconsistent", Burke, Wasserman, and Sim et al.[1] give an overview of the HF-DFT method, also known as the density-corrected DFT (DC-DFT) method. The essential stratagem of HF-DFT actually goes back to the dawn of molecular DFT, where—as a pragmatic expedient that permitted quickly retrofitting DFT into a wave function *ab initio* program system—new GGA or mGGA exchange–correlation (XC) functionals would be evaluated for densities generated from pure Hartree–Fock self-consistent field (SCF) orbitals.[2,3] (A related practice, "post-local-density approximation (LDA) DFT", consisted of using LDA densities from early DFT codes in the same fashion,[4] so a new GGA or meta-GGA functional could quickly be assessed before going to the trouble of a self-consistent implementation).

Pure HF orbitals are rigorously free of self-interaction errors (SIEs). While, for instance, PBE evaluated using HF orbitals will not be SIE-free, Lonsdale and Goerigk observed[5] that HF-DFT functionals systematically have a lower SIE than the corresponding self-consistent functional.

More generally speaking, HF-XC (where XC is a given exchange–correlation functional) arguably might be beneficial in any situation where the chief source of error in XC is a misshapen density, rather than intrinsic exchange and correlation errors.

Very recently, Burke and co-workers found[6,7] that HF-DFT is beneficial in the treatment of halogen and pnictogen bonds. As we have some experience in halogen bonding (e.g.,[8,9]), we were intrigued by this finding. We were also motivated in part by our work on minimally empirical double hybrids[10,11] and by the question whether HF-DFT would still be beneficial at the high percentages of Hartree–Fock exchange such functionals typically entail.

It then occurred to us that, to our knowledge, no evaluation had yet been carried out of HF-DFT with a large and chemically diverse benchmark suite like GMTKN55 (general main-group thermochemistry, kinetics, and noncovalent interactions—55 problem types[12]). We present such an analysis below for several hybrid GGA and meta-GGA

A

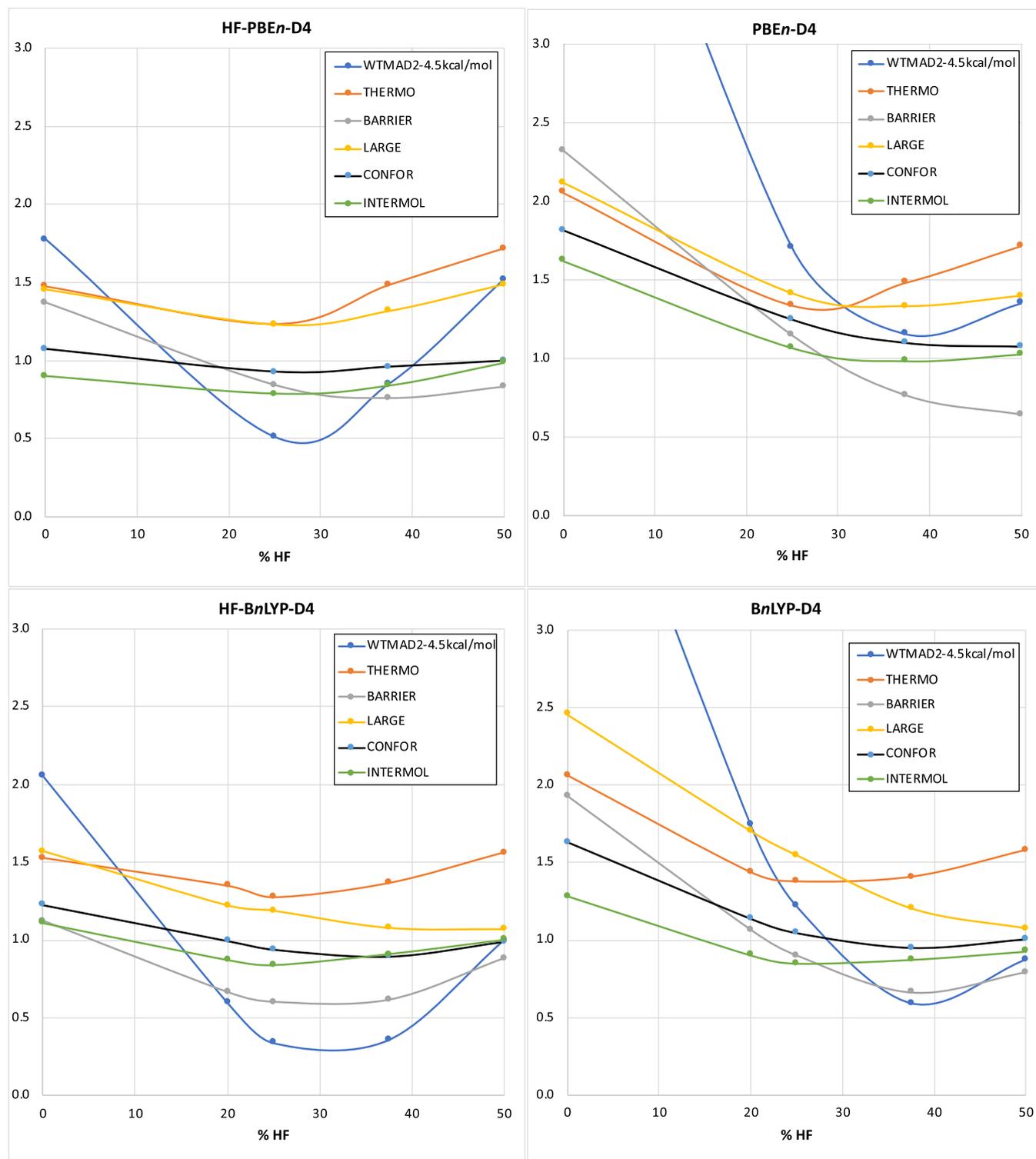

**Figure 1.** Dependence of WTMAD2 (kcal/mol) and of the five top-level subsets on the percentage of HF exchange for HF-PBE$n$-D4, PBE$n$-D4, HF-B$n$LYP-D4, and B$n$LYP-D4.

sequences, with percentages of Hartree−Fock exchange in the functional varying from 0 to 50%.

## COMPUTATIONAL DETAILS

The GMTKN55 benchmark suite of Goerigk, Grimme, and co-workers[12] was used as our training set. This data set consists of nearly 1500 energy differences entailing almost 2500 energy evaluations. Its 55 subsets can be conveniently grouped into 5 classes: thermochemistry, barrier heights, intermolecular (non-covalent) interactions, conformers (dominated by intra-molecular noncovalent interactions), and reaction energies for large systems. A detailed description of all 55 subsets can be found in the original paper[12] and a brief itemized summary in the present paper's Supporting Information.

As our primary error metric, we used WTMAD2 (weighted mean absolute deviation, type 2, as defined in the original GMTKN55 paper[12]): its expression has the form

$$\text{WTMAD2} = \frac{1}{\sum_{i=1}^{55} N_i} \cdot \sum_{i=1}^{55} N_i \cdot \frac{56.84 \text{ kcal/mol}}{|\overline{\Delta E}|_i} \cdot \text{MAD}_i$$

where $|\overline{\Delta E}|_i$ are the mean absolute values of all the reference energies for subset $i$ = 1 to 55 (thus "normalizing" errors of all subsets to the same scale, so to speak), $N_i$ is the number of systems in subset $i$, and $\text{MAD}_i$ represents the mean absolute difference between our calculated and the original reference energies for subset $i$. We also considered the WTMAD2 contributions for the five primary categories as well as for the individual subsets.

The primary reason, in ref 12 and in the present work, to choose an MAD-based metric over a root-mean-square-deviation (rmsd)-based one is that MAD is more "robust" in the statistical sense[13] (e.g., less prone to strongly vary because of one or two outliers). The rmsd/MAD ratio for a normal distribution should be[14] $(\pi/2)^{1/2}$ = 1.253314... ≈ 5/4; an abnormally large or small ratio is almost invariably an indicator for outliers. Hence, we monitored the rmsd/MAD ratio for each subset throughout the work.

Our first HF-DFT explorations (involving PBE hybrids) were carried out using the Gaussian16 package[15] and the remainder using ORCA 4.2.1,[16] with all running on the Faculty of Chemistry's CHEMFARM high-performance computing facility. Self-consistent PBE[17] and PBE0[18] calculations were carried out using Q-CHEM.[19]

Reference geometries from ref 12 were used 'as is', without any further optimization. For most of the systems in HF-DFT and SC-DFT single-point electronic structure calculations, the Weigend−Ahlrichs def2-QZVPP[20] basis set was used, except for the five anion-containing subsets WATER27, RG18, IL16, G21EA, and AHB21 where we used diffuse-function augmented def2-QZVPPD.[21] Density fitting for the Coulomb and exchange part was used throughout in ORCA, in conjunction with the appropriate def2/JK density fitting basis sets.[22] In the ORCA calculations, we employed GRID 5 as the integration grid, except for the SCAN (strongly constrained and appropriately normed[23] [nonempirical] meta-GGA functional) and HF-SCAN series, where we used the larger GRID 6 because of SCAN's well-documented[24] strong integration grid sensitivity. TightSCF convergence criteria were used throughout. In Q-Chem, we used the SG-3 grid[25] throughout sample HF-DFT inputs for Gaussian 16 and ORCA can be found in the Supporting Information).

One series of calculations, HF-PBE$n$, consisted of HF-DFT counterparts of PBE[17] (0% HF exchange), PBE0[18] (25% HF exchange), PBE38[26] (37.5% HF exchange), and PBE50[27] (50% HF exchange) as well as the respective self-consistent functionals for comparison. A second was HF-B$n$LYP, consisting of HF-DFT versions of BLYP[28,29] (0% HF exchange), B20LYP (20% HF exchange), B1LYP[30] (25% HF exchange), B38LYP (37.5% HF exchange), and BHLYP[29,31] (50% HF exchange), again compared with their self-consistent variants (we note in passing that the widely used B3LYP,[31] unlike B20LYP, uses a mix of 8% Slater LSDA and 72% Becke88 exchange[28] and a mix of 19% VWN5 LSDA correlation[32] and 81% LYP GGA correlation.[29] Also, for the avoidance of doubt, BHLYP refers to 50% Becke88, 50% HF exchange, and 100% LYP correlation rather than the B3LYP-inspired LDA-GGA mix of Shao, Head-Gordon, and Krylov[33] implemented in Q-CHEM).

We briefly discuss a third series comprising HF-DFT and self-consistent DFT versions of meta-GGA TPSS[34] (0% HF exchange), TPSSh (10% HF exchange),[35] TPSS0 (25% HF exchange), TPSS38 (37.5% HF exchange), and TPSS50 (50% HF exchange). Finally, we explore HF-DFT and self-consistent series of the recent SCAN meta-GGA functional:[23] SCAN (0% HF exchange), SCAN10 (10% HF exchange), SCAN0 (25% HF exchange), SCAN38 (37.5% HF exchange), and SCAN50 (50% HF exchange).

In order to treat dispersion on an equal footing everywhere, the recent DFT-D4 model,[36,37] as implemented in Grimme's standalone dftd4 program (https://www.chemie.uni-bonn.de/pctc/mulliken-center/software/dftd4), was employed throughout. Since not for all self-consistent cases, "official" parameters were available and none at all were available for HF-DFT-D4, we refitted the three nontrivial parameters $s_8$, $a_1$, and $a_2$ for each functional by minimizing WTMAD2 over GMTKN55 (as we do not consider double hybrids[11,38] in the present work, the fourth parameter is constrained to be $s_6$ = 1 throughout, as is the prefactor for the three-body Axilrod−Teller−Muto[39,40] correction, $c_{\text{ATM}}$ = 1). The D4 parameter optimizations were performed using Powell's derivative-free constrained optimizer, BOBYQA (Bound Optimization BY Quadratic Approximation)[41] and a collection of scripts developed in-house.

All parameter values and the corresponding WTMAD2s and five-component breakdowns of the same can be found in Table S2 in the Supporting Information, where the corresponding data for any available "official" parameterizations are also given with proper references.

## RESULTS AND DISCUSSION

**GGA Series: PBE$n$-D4 versus HF-PBE$n$-D4 and B$n$LYP-D4 versus HF-B$n$LYP.** Our discussion will focus mostly on the PBE$n$-D4 versus HF-PBE$n$-D4 series, but the behavior of the B$n$LYP-D4 versus HF-B$n$LYP-D4 series (see the Supporting Information) is, by and large, quite similar. In Figure 1, we summarize for all four scenarios the dependence on the percentage of HF exchange of WTMAD2 as well as its five top-level subdivisions: basic thermochemistry (THERMO), reaction barrier heights (BARRIER), large-molecule reactions including isomerizations (LARGE), conformational equilibria (CONFOR), which are generally driven by intramolecular noncovalent interactions, and intermolecular interactions (INTER).

Intriguingly, for both PBE$n$-D4 and B$n$LYP-D4, the overall minimum is not, as one might expect, near the 25% advocated for thermochemistry in PBE0[42] on the basis of a perturbation theoretical argument[43] but near 37.5% or 3/8, consistent with the PBE38 functional proposed by Grimme and co-workers[26,44] and akin to the earlier mPW1K (modified Perdew−Wang[45] with one parameter for kinetics) functional with 42.8% HF exchange.[46] The minor loss in accuracy for basic thermochemistry is more than compensated by the improvements for barrier heights and for large-molecule isomerization reactions. In contrast, HF-PBE$n$-D4 and HF-B$n$LYP-D4 see WTMAD2 minima at lower percentages of HF exchange, closer to 25% or 1/4 in the PBE case and to 30% in the BLYP case. Barrier heights are the reason behind this optimum % HF shift from self-consistent to HF-PBE$n$-D4; small-molecule thermochemistry shows minima near 25% for *both* PBE$n$-D4 and HF-PBE$n$-D4.

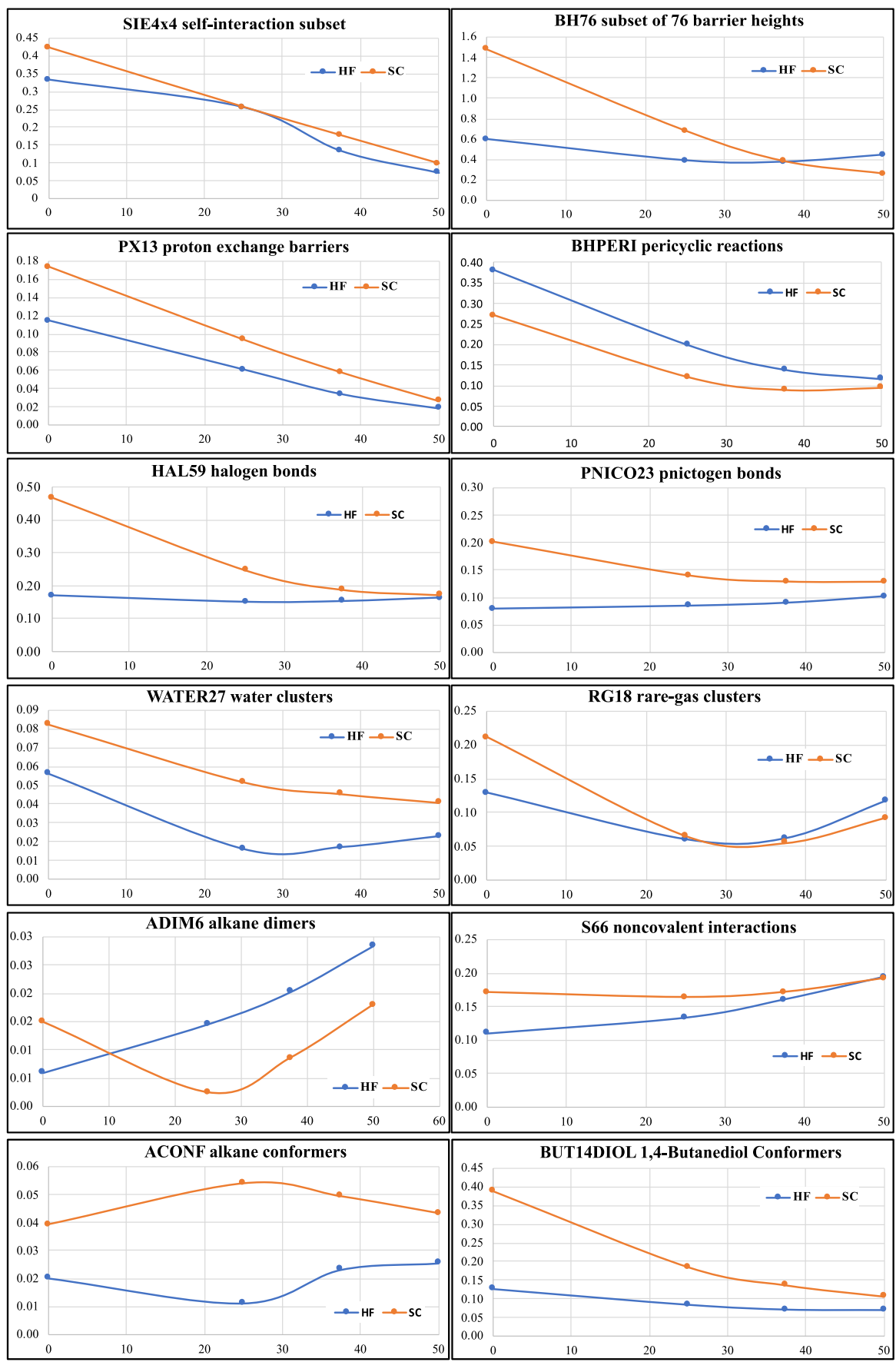

**Figure 2.** continued

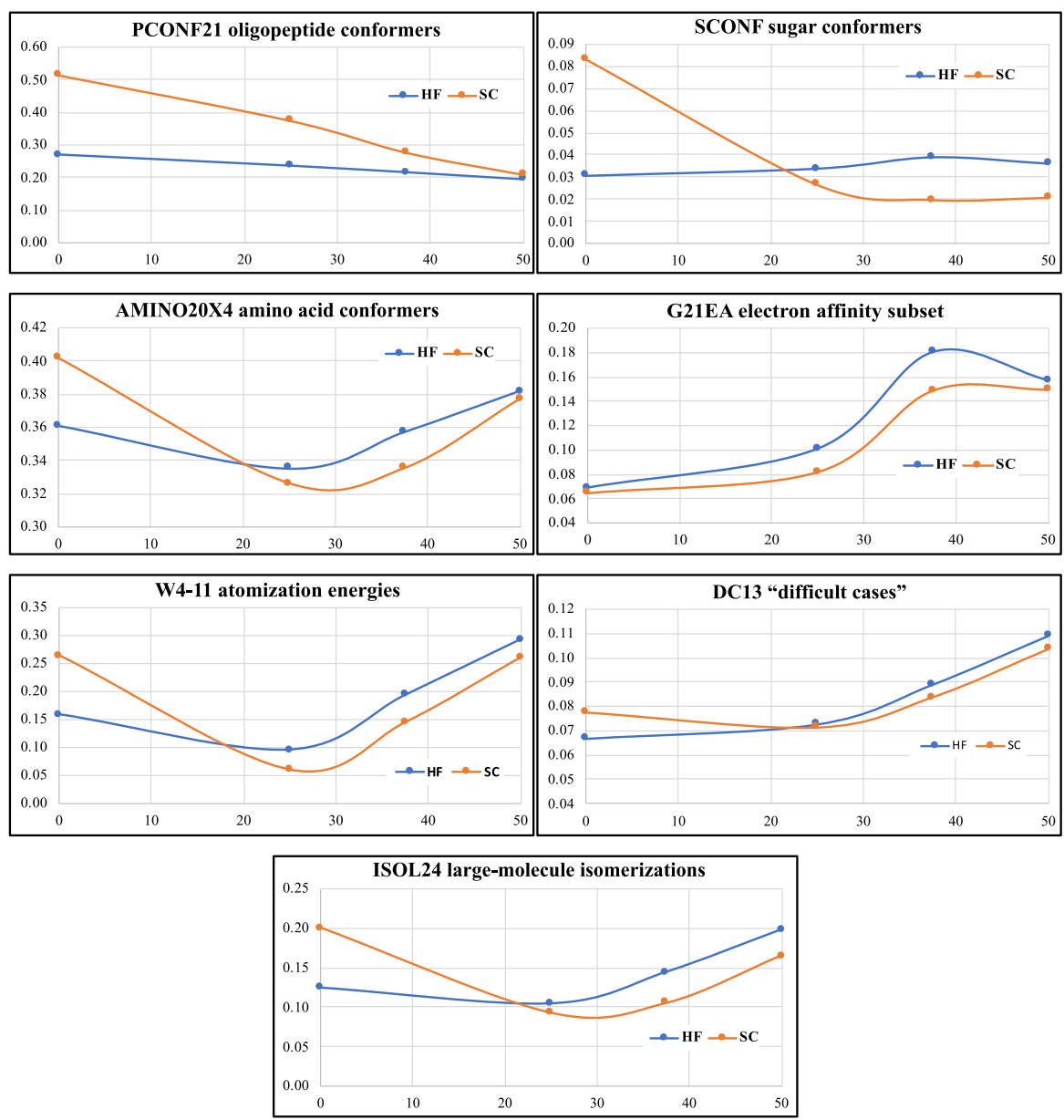

**Figure 2.** Dependence on the percentage of HF exchange for self-consistent PBE$n$-D4 (SC) and HF-PBE$n$-D4 (HF) of the WTMAD2 (kcal/mol) contributions for the individual subsets SIE4x4, BH76, PX13, BHPERI, HAL59, PNICO23, WATER27, RG18, ADIM6, S66, alkane conformers (ACONF), 1,4-butanediol conformers (BUT14DIOL), oligopeptide conformers (PCONF21), sugar conformers (SCONF), amino acid conformers (AMINO20X4), G21EA, W4-11, DC13, and large-molecule isomerization(ISOL24) subsets. A similar figure for the B$n$LYP-D4 and HF-B$n$LYP-D4 cases appears as Figure S1 in the Supporting Information.

It has been well known for over 2 decades[46−48] that barrier heights of radical reactions are systematically underestimated by GGAs and that hybrids with high percentages of HF exchange (e.g., refs [48,49]) yield the best performance. This has been ascribed[50−52] to SIE, which is reduced as the percentage of HF exchange is increased, and it has been shown convincingly (e.g., refs [52,53]) that self-interaction corrections (and to a lesser extent, HF-DFT[54]) improve barrier heights predicted both by GGAs and by meta-GGAs (the fact that meta-GGAs have advantages over GGAs for barrier heights, quite aside from this issue, was shown by Zheng et al.[55]). In some situations, however, such as pnictogen inversion barriers, HF exchange counterintuitively lowers barriers,[56] which has been rationalized by Mahler et al.[57] as occurring in situations where bond orbitals in the transition state have more $s$ and less $p$ character than in reactants and products. However, the reader should also see the work of Truhlar and co-workers showing additional complexity involving multireference character in the process.[58]

A reviewer enquired about "optimal" %HF for the total atomization energies (TAEs) of given molecules and how they are distributed. In ref [59], we showed for the W4-11 data set[60] that the TAEs depend almost perfectly negative-linearly on % HF and that the relative slopes can be exploited for a nondynamical correlation diagnostic. Hence, if a given DFT functional overestimates the reference TAE, the "optimal" % HF value for that molecule will go up to compensate (and conversely). Therefore, unless there is very little spread in the errors of the functional (i.e., mostly systematic errors), individual "optimal" %HF values will run the gamut.

Table 1. MAD and MSD (Mean Absolute and Mean Signed Deviations, kcal/mol) of RG18, S66, and Four Subcategories of S66 for PBE*n*, HF-PBE*n*, B*n*LYP, and HF-B*n*LYP with and without D4 Dispersion

| functionals | hydrogen bond systems 1−23 MAD | MSD | π-stacking systems 24−33 MAD | MSD | London dispersion systems 34−46 MAD | MSD | mixed-influence systems 47−66 MAD | MSD | full S66 MAD | MSD | RG18 MAD | MSD |
|---|---|---|---|---|---|---|---|---|---|---|---|---|
| HF-PBE-D4 | 0.20 | 0.09 | 0.44 | 0.44 | 0.18 | −0.05 | 0.22 | 0.17 | 0.24 | 0.14 | 0.11 | 0.11 |
| PBE-D4 | 0.59 | 0.59 | 0.40 | −0.07 | 0.31 | 0.15 | 0.15 | 0.09 | 0.37 | 0.25 | 0.18 | 0.16 |
| HF-PBE0-D4 | 0.34 | 0.34 | 0.32 | 0.32 | 0.22 | −0.18 | 0.26 | 0.19 | 0.29 | 0.19 | 0.05 | 0.01 |
| PBE0-D4 | 0.60 | 0.59 | 0.26 | −0.04 | 0.31 | 0.09 | 0.21 | 0.13 | 0.37 | 0.26 | 0.06 | 0.03 |
| HF-PBE38-D4 | 0.48 | 0.48 | 0.29 | 0.29 | 0.25 | −0.21 | 0.29 | 0.22 | 0.35 | 0.24 | 0.05 | −0.02 |
| PBE38-D4 | 0.69 | 0.69 | 0.14 | 0.05 | 0.18 | −0.05 | 0.25 | 0.19 | 0.37 | 0.30 | 0.05 | 0.00 |
| HF-PBE50-D4 | 0.67 | 0.67 | 0.20 | 0.20 | 0.31 | −0.30 | 0.33 | 0.26 | 0.42 | 0.28 | 0.10 | −0.09 |
| PBE50-D4 | 0.78 | 0.78 | 0.10 | 0.07 | 0.19 | −0.15 | 0.29 | 0.23 | 0.42 | 0.32 | 0.08 | −0.06 |
| HF-PBE | 1.65 | −1.65 | 4.39 | −4.39 | 3.80 | −3.80 | 2.30 | −2.30 | 2.69 | −2.69 | 0.41 | −0.41 |
| PBE | 0.74 | −0.62 | 4.10 | −4.10 | 3.24 | −3.24 | 1.92 | −1.92 | 2.10 | −2.06 | 0.27 | −0.22 |
| HF-PBE0 | 1.13 | −1.13 | 4.03 | −4.03 | 3.54 | −3.54 | 1.99 | −1.99 | 2.30 | −2.30 | 0.46 | −0.46 |
| PBE0 | 0.64 | −0.55 | 3.98 | −3.98 | 3.24 | −3.24 | 1.81 | −1.81 | 2.01 | −1.98 | 0.36 | −0.36 |
| HF-BLYP-D4 | 0.22 | −0.20 | 0.15 | 0.09 | 0.76 | −0.76 | 0.35 | −0.34 | 0.36 | −0.31 | 0.32 | −0.32 |
| BLYP-D4 | 0.20 | −0.04 | 0.48 | 0.42 | 0.41 | −0.41 | 0.35 | −0.34 | 0.33 | −0.13 | 0.30 | −0.29 |
| HF-B20LYP-D4 | 0.29 | 0.29 | 0.16 | 0.01 | 0.60 | −0.60 | 0.20 | −0.11 | 0.30 | −0.05 | 0.22 | −0.22 |
| B20LYP-D4 | 0.40 | 0.39 | 0.27 | 0.27 | 0.37 | −0.37 | 0.18 | −0.11 | 0.31 | 0.07 | 0.19 | −0.18 |
| HF-B1LYP-D4 | 0.33 | 0.33 | 0.15 | 0.01 | 0.57 | −0.57 | 0.19 | −0.10 | 0.31 | −0.03 | 0.20 | −0.20 |
| B1LYP-D4 | 0.49 | 0.49 | 0.21 | 0.21 | 0.36 | −0.36 | 0.16 | −0.07 | 0.32 | 0.11 | 0.17 | −0.15 |
| HF-B38LYP-D4 | 0.69 | 0.69 | 0.24 | −0.08 | 0.47 | −0.47 | 0.19 | 0.06 | 0.43 | 0.15 | 0.16 | −0.13 |
| B38LYP-D4 | 0.83 | 0.83 | 0.18 | 0.15 | 0.27 | −0.27 | 0.20 | 0.14 | 0.43 | 0.30 | 0.12 | −0.08 |
| HF-BHLYP-D4 | 1.00 | 1.00 | 0.26 | 0.07 | 0.24 | −0.24 | 0.28 | 0.26 | 0.52 | 0.39 | 0.10 | −0.05 |
| BHLYP-D4 | 1.06 | 1.06 | 0.24 | 0.10 | 0.25 | −0.10 | 0.27 | 0.25 | 0.54 | 0.44 | 0.08 | −0.02 |

The gain from using HF orbitals is the greatest for *n* = 0, namely, 3.64 kcal/mol for HF-PBE-D4 and 2.81 kcal/mol for HF-BLYP-D4. It is still significant for HF-PBE0-D4, 1.19 kcal/mol, and HF-B1LYP-D4, 0.88 kcal/mol. Then, it continues to decay monotonically until it becomes a net liability for HF-BHLYP-D4 and HF-PBE50-D4.

Breaking down by the five top-level components, we see that HF-PBE*n*-D4 nearly "flattens the curves", compared to PBE*n*-D4, for conformers and intermolecular interactions. In the low-HF region, barrier heights benefit most significantly, but this vanishes around HF-PBE38-D4: self-consistent densities with high percentages of HF exchange still do better.

Benefits are seen at zero to moderate HF exchange for large-molecule reactions and for basic thermochemistry.

For the B*n*LYP series, the curves for conformers and intermolecular interactions are flatter to begin with, presumably because LYP is a very short-ranged correlation functional and dispersion corrections play a much larger role here (by way of illustration: for the simple D2 empirical dispersion correction, which just has a simple scaling factor $s_6$ as an empirical parameter, $s_6$ = 0.65 for PBE0 and 1.2 for B3LYP[61]). Otherwise, things are much the same as those for the PBE series: HF-DFT is definitely beneficial at zero to moderate (about 25%) HF exchange and becomes detrimental overall at 50% HF exchange.

Let us now zoom in on individual subsets that vary the most (Figure 2). First, the SIE4x4 self-interaction subset benefits for all percentages of HF exchange, reducing it by 10−25%.

The BH76 subset of 76 barrier heights is the set union of Truhlar's HTBH38 and NHTBH38 hydrogen- and non-hydrogen-transfer barrier heights.[62,63] Here, the improvement granted by HF-DFT is quite dramatic at zero or low HF exchange—at 50%, HF-DFT is actually detrimental, while at 37.5%, it is a wash. For the PX13 proton-exchange barriers, we again see a substantial benefit at 0% and a significant one at 25%. Intriguingly, for the BHPERI pericyclic reactions, HF-DFT appears to be detrimental, but this may be masked by the known fact[64] that dispersion contributions, which disproportionately stabilize the transition states, are quite significant for these barriers.

For the HAL59 halogen bonds, the PNICO23 pnictogen bonds, and the WATER27 water clusters (multiple hydrogen bonds), HF-PBE*n*-D4 is beneficial across the board, while for the B*n*LYP series, where the contribution of the D4 dispersion correction is larger, we actually find that HF-B*n*LYP-D4 is detrimental for WATER27 but beneficial for higher percentages of HF exchange: B*n*LYP-D4 underbinds on average for 0% but overbinds for the other, and HF-B*n*LYP lowers the interaction energies.

From a symmetry-adapted perturbation theory (SAPT) point of view,[65] the three sets above have in common that they are predominantly driven by electrostatic interactions rather than dispersion. Therefore, what about dispersion-dominant subsets? The RG18 rare-gas clusters and the ADIM6 *n*-alkane dimers are archetypical examples: in neither is HF-DFT very helpful, which makes sense since the interactions occur at a longer distance.

The S66 noncovalent interaction benchmark[66] (see ref 67 for a recent revision) is a more mixed bag. Breaking down by subsets reveals the answer here: systems 1−23 are hydrogen bonds, systems 24 through 33 are π-stacking, systems 34 through 46 are London dispersion complexes, and the remainder are mixed-influence. Again, we see that HF-DFT is beneficial mostly for the hydrogen bond component: this is most pronounced for the PBE series (Table 1).

Comparison of the HF-PBE and PBE series without dispersion correction (i.e., without the "confounding factor" of different optimized dispersion parameters) shows that RG18, ADIM6, and the London dispersion subset of S66 (i.e., systems 34−46) are all more weakly bound in HF-PBEn than in PBEn, with the gap narrowing as n increases. Upon introducing a dispersion correction, the situation may arise (e.g., for the BLYP-D4 series) where DFT-D4 is already underbound, in which case HF-DFT-D4 will make things worse, or DFT-D4 may be overbound and HF-DFT-D4 sets things to rights (as seen in Table 1 for HF-PBE-D4 and HF-PBE0-D4).

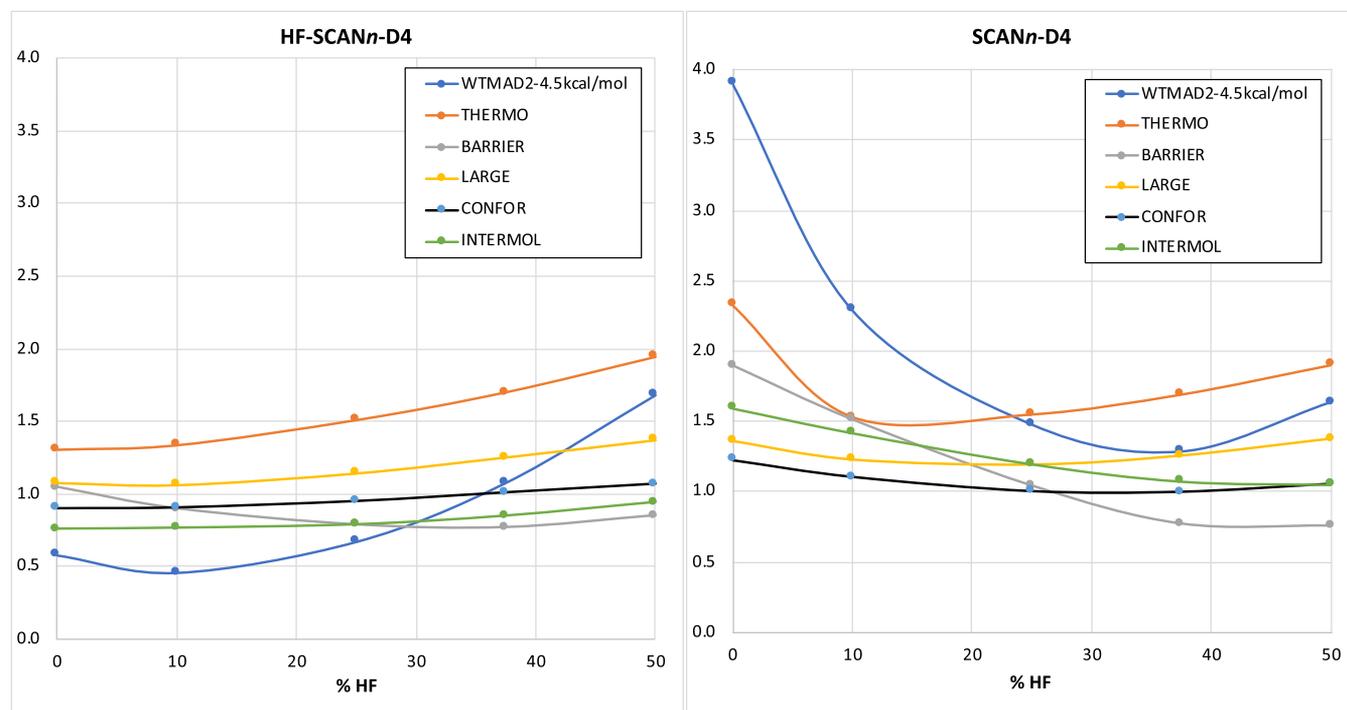

**Figure 3.** Dependence of WTMAD2 (kcal/mol) and the five top-level subsets on the percentage of HF exchange for HF-SCANn-D4 and SCANn-D4.

What is the effect of the HF density here really? We attempted to create a difference density plot between HF-PBE and PBE for the water dimer, but nothing of note is easily visible. However, the long-distance tail of the HF density clearly exhibits the expected[68] exponential decay, while the self-consistent PBE density decays much more slowly: an illustrative plot for the Ar atom can be found in the Supporting Information.

As for π-stacking, it entails features like quadrupole−quadrupole interactions that are difficult to describe for hybrid GGAs with or without dispersion correction, as discussed at length in ref 69. Yet in Table 1, the HF-BLYP-D4 series is seen to perform quite well; in contrast, fully self-consistent PBEn-D4 with a large n clearly does better than HF-PBEn-D4.

Turning now to conformers: in light of the above, it is not surprising that a series like the 1,4-butanediol conformers,[70] the conformer equilibria of which are dominated by the making and breaking of internal hydrogen bonds, would benefit from HF-DFT. So do sugar and oligopeptide conformers at the low-HF exchange end. Nevertheless, alkane conformers also benefit slightly. Amino acid conformers[71,72] are a mixed bag as the equilibria for residues with hydrophobic side chains (e.g., valine, isoleucine, and leucine) will be driven more by dispersion, and those for residues like lysine and arginine will be driven more by electrostatics.

In principle, electron affinities should get better at least for pure BLYP and PBE because in HF-PBE, the anion HOMO is at least bound. In practice, Tschumper and Schaefer[73] already showed back in 1997 that because of spatial confinement by the finite basis set, even BLYP and PBE EAs are fairly reasonable (see also de Oliveira et al.[74] for more discussion). In the present work, we see no improvement in actual EAs from using HF-DFT, and for hybrid functionals, HF-DFT actually appears to do more harm than good.

For the W4-11 set of atomization energies,[60] HF-BLYP-D4 and HF-PBE-D4 are definitely superior over the pure-DFT BLYP-D4 and PBE-D4 functionals, respectively, but with some HF exchange introduced at the orbital stage, self-consistency appears to be more beneficial. Similarly, for the DC13 "difficult cases" benchmark, HF-BLYP-D4 and HF-PBE-D4 are helpful, but here, a benefit is seen all along the BnLYP series. Intriguingly, the large-molecule isomerizations exhibit the same behavior.

**meta-GGA Series, Particularly HF-SCAN$n$-D4 versus SCAN$n$-D4.** However, perhaps the above behaviors might not be replicated for meta-GGAs such as TPSS. Therefore, we carried out a similar investigation for the HF-TPSS$n$-D4 and TPSS$n$-D4 series. Detailed results can be found in the Supporting Information, but the bottom line is that HF-TPSS$n$-D4 behaves quite similarly to the HF-PBE$n$-D4 and HF-B$n$LYP-D4 series.

The recent nonempirical SCAN functional, however, departs from this pattern to some degree (Figure 3). The graph for self-consistent SCAN$n$-D4 qualitatively appears fairly similar to its counterparts for PBE$n$-D4, B$n$LYP-D4, and TPSS$n$-D4: it even has the same overall minimum at or near 37.5% (3/8) of HF exchange.

However, when HF orbitals are used, the minimum shifts not to 25% (1/4) as for the abovementioned series but to about 10%. What is more, HF-SCAN-D4 without any "exact" exchange in the post-HF step performs nearly as well as the "optimum" HF-SCAN10-D4. Comparing this to the compilation of WTMAD2 values for other DFT functionals[11,75] (or with slightly different basis set choices, in refs[12,76]), the WTMAD2 of 5.08 kcal/mol is superior to all hybrid GGAs and meta-GGAs except for combinatorially parameterized range-separated hybrids like ωB97X-V (WTMAD2 = 3.96 kcal/mol) and ωB97M-V (WTMAD2 = 3.26 kcal/mol).[77,78] It must be emphasized that HF-SCAN-D4 has no empirical parameters in the electronic structure part, just the three parameters of the D4 dispersion correction (the fitted D4 parameters can be found in Table S2 in the Supporting Information). It will be more attractive in terms of computational cost than other hybrid meta-GGAs since the cost will basically be that of a simple HF calculation, followed by a single evaluation of the SCAN meta-GGA functional.

Table 2. MAD (kcal/mol) and MSD (kcal/mol) Values of Four Subcategories and the Full S66 Set and RG18 Subset for HF-SCANn-D4 and SCANn-D4

| functionals | hydrogen bond systems 1−23 | | π-stacking systems 24−33 | | London dispersion systems 34−46 | | mixed-influence systems 47−66 | | full S66 | | RG18 | |
|---|---|---|---|---|---|---|---|---|---|---|---|---|
| | MAD | MSD | MAD | MSD | MAD | MSD | MAD | MSD | MAD | MSD | MAD | MSD |
| HF-SCAN-D4 | 0.21 | 0.09 | 0.57 | 0.57 | 0.47 | −0.45 | 0.23 | 0.02 | 0.32 | 0.03 | 0.05 | 0.01 |
| SCAN-D4 | 0.73 | 0.73 | 0.10 | −0.03 | 0.34 | −0.34 | 0.23 | 0.01 | 0.41 | 0.19 | 0.14 | 0.09 |
| HF-SCAN10-D4 | 0.31 | 0.24 | 0.44 | 0.44 | 0.44 | −0.42 | 0.24 | 0.07 | 0.33 | 0.09 | 0.06 | −0.01 |
| SCAN10-D4 | 0.79 | 0.79 | 0.08 | −0.01 | 0.23 | −0.22 | 0.20 | 0.10 | 0.39 | 0.26 | 0.11 | 0.06 |
| HF-SCAN0-D4 | 0.45 | 0.42 | 0.26 | 0.26 | 0.41 | −0.41 | 0.25 | 0.11 | 0.35 | 0.14 | 0.06 | −0.04 |
| SCAN0-D4 | 0.84 | 0.84 | 0.08 | −0.02 | 0.16 | −0.14 | 0.23 | 0.18 | 0.41 | 0.32 | 0.08 | 0.01 |
| HF-SCAN38-D4 | 0.59 | 0.58 | 0.15 | 0.15 | 0.37 | −0.37 | 0.28 | 0.16 | 0.39 | 0.20 | 0.07 | −0.05 |
| SCAN38-D4 | 0.89 | 0.89 | 0.10 | −0.05 | 0.14 | −0.11 | 0.27 | 0.22 | 0.43 | 0.35 | 0.06 | −0.02 |
| HF-SCAN50-D4 | 0.79 | 0.79 | 0.13 | 0.13 | 0.28 | −0.26 | 0.32 | 0.26 | 0.45 | 0.32 | 0.08 | −0.06 |
| SCAN50-D4 | 0.98 | 0.98 | 0.10 | −0.03 | 0.11 | −0.06 | 0.32 | 0.29 | 0.48 | 0.42 | 0.06 | −0.04 |
| HF-SCAN | 0.51 | −0.45 | 1.38 | −1.38 | 2.06 | −2.06 | 0.94 | −0.94 | 1.08 | −1.06 | 0.20 | −0.20 |
| SC-SCAN | 0.57 | 0.40 | 1.24 | −1.24 | 1.45 | −1.45 | 0.71 | −0.62 | 0.89 | −0.53 | 0.22 | −0.03 |
| HF-SCAN0 | 0.39 | −0.15 | 1.75 | −1.75 | 2.08 | −2.08 | 0.91 | −0.88 | 1.09 | −0.99 | 0.25 | −0.25 |
| SC-SCAN0 | 0.57 | 0.36 | 1.73 | −1.73 | 1.72 | −1.72 | 0.80 | −0.72 | 1.04 | −0.70 | 0.19 | −0.16 |

How does HF-SCAN achieve this quite remarkable result? If we look at the five top-level subclasses of GMTKN55, we find that HF-SCANn-D4 nearly flattens the intermolecular interaction curve and improves across the board, especially for lower HF fractions. For conformer energy differences (mainly driven by intramolecular noncovalent interactions), HF-SCANn-D4 flattens and improves at low HF exchange, while neither helping nor harming much at high HF exchange. By and large, the same is observed for large-molecule reaction energies. For small-molecule thermochemistry, a noticeable improvement is seen at low HF exchange, leading to HF-SCAN-D4 indeed becoming the optimum for that subset. Only for barrier heights is a minimum still found at a high fraction of exact exchange (at or near HF-SCAN38-D4), but the curve is much flatter than that for the SCANn-D4 series and indeed for the other HF-functional-D4 series we have considered above.

Now, if we focus on individual subsets (Figure S2 in the Supporting Information), for the SIE4x4 self-interaction subset, benefits are seen up to 50% HF exchange, beyond which self-consistent SCANn-D4 and HF-SCANn-D4 perform similarly. BH76, BHPERI, and PX13 exhibit similar trends to the other tested GGA and meta-GGA cases.

In the cases of HAL59, PNICO23, and WATER27, HF variants perform better than the self-consistent counterparts, with (as expected) the performance gap monotonically decreasing as HF exchange is progressively introduced into the DFT part.

For the rare-gas interaction subset RG18, HF-SCAN-D4 is quite beneficial over SCAN-D4, but as HF exchange is introduced in the DFT parts, the self-consistent variants "catch up" so that already HF-SCAN0-D4 and SCAN0-D4 have similarly low error statistics. In contrast, for the n-alkane dimer interaction subset ADIM6, which should similarly be driven mostly by London dispersion, self-consistent SCANn-D4 is consistently superior to HF-SCANn-D4, although both improve as $n$ is increased (unlike the behavior for the PBE series).

For the noncovalent interaction subset S66, HF-SCANn-D4 outperforms its self-consistent counterpart across the range. Breakdown by subsets reveals that this is entirely driven by the hydrogen bonds; the π-stacks and London subsets actually show deterioration when using HF rather than self-consistent densities, while for the mixed-influence complexes, it does not appear to matter which (Table 2).

1,4-butanediol conformers BUT14DIOL systematically benefit from HF densities, although the gap with self-consistent SCANn-D4 narrows as the percentage $n$ of HF exchange is increased. So do sugar conformers SCONF at low HF exchange, while at SCAN0-D4 and beyond, self-consistent densities are preferred. Peptide conformers PCONF and amino acid conformers AMINO20X4 do not benefit from HF-DFT, whereas alkane conformers benefit insignificantly beyond 40%.

In contrast, for large-molecule isomerization reactions (ISOL24), one can see a benefit to HF-SCAN-D4 over SCAN-D4 and more marginally for HF-SCAN10-D4 over SCAN10-D4, but beyond this point, using HF densities is no longer beneficial.

Turning now to three subsets with some nondynamical correlation effects: for the W4-11 atomization energies (note: reference values[60] are approximate FCI/CBS rather than CCSD(T)/CBS), HF-SCAN-D4 clearly outperforms SCAN-D4, but for all hybrids considered, HF-SCANn-D4 does more poorly than SCANn-D4. No benefit from HF density is seen for the 13 difficult case (DC13) subset. Here too, for the G21EA electron affinity subset, HF-DFT does not help at all, as observed for the other three cases (PBE$n$, B$n$LYP, and TPSS$n$).

**Additional Remarks.** Finally, what about range-separated hybrids? We carried out limited testing with the CAM-B3LYP[79] and LC-ωPBE[80] range-separated hybrids. To cut a long story short, it appears that range-separated hybrids do not benefit much from HF-DFT as so much HF exchange is already present at a long range.

Double-hybrid functionals typically entail percentages of HF exchange ranging from 50% for PBE0-DH[81] and 53% for B2PLYP[38] to $(1/2)^{1/3} \approx 79.3\%$ for PBE0-2[82] and 81% for B2NC-PLYP,[83] with the best performers situated in the range of 62.2% for ωB97M(2)[84] via 65% for B2GP-PLYP[85] to 69% for revDSD-PBEP86.[10] In view of the preceding, it is clear that all these percentages are well beyond the range where DC-DFT would be beneficial in any way.

# ■ CONCLUSIONS

From our comparative evaluation against the GMTKN55 benchmark suite of the PBE$n$-D4 and HF-PBE$n$-D4 series as well as the corresponding series involving B$n$LYP, TPSS$n$, and SCAN$n$, we were able to draw the following conclusions:

(a) For the self-consistent series, the global minimum in terms of WTMAD2 lies near 3/8 (or 37.5%) of HF exchange, that is, the PBE38-D4 functional proposed by Grimme;[26,44] the loss of accuracy in small-molecule thermochemistry is compensated by gains in accuracy elsewhere, notably barrier heights. In contrast, for HF-PBE$n$-D4 and HF-B$n$LYP-D4, the global minimum lies closer to 1/4 (25%) of HF exchange. The drivers for this shift are barrier heights: while the region near 25% is the "thermochemical comfort zone" for both HF-PBE$n$-D4 and PBE$n$-D4, barrier heights entail too large an error at 25% for PBE0-D4, which is greatly reduced in HF-PBE0-D4.

(b) In general, the benefits of HF-DFT are the greatest for pure GGAs and decrease monotonically with increasing HF exchange: for 50% and above, self-consistent functionals are actually superior to HF-DFT, implying that HF-DFT does not represent a general route to improve double hybrids. Similarly, HF-DFT does not offer a route for further improvement in range-separated hybrids.

(c) With moderate HF exchange (e.g., HF-PBE0-D4 and HF-B1LYP-D4), the benefits of HF-DFT are the greatest in noncovalent interactions that from a SAPT perspective have a strong electrostatic component, such as hydrogen bonds, halogen bonds, pnictogen bonds, and the like. These benefits are also seen in conformers and isomer equilibria that are primarily driven by intramolecular hydrogen bonding (and the like). For London dispersion-dominated problems like rare-gas clusters and alkane dimers, HF is beneficial if the underlying functional is overbinding, and detrimental otherwise.benefit from HF-DFT if the underlying functional is overbinding dispersion problems, do not benefit, while HF-DFT does more harm than good for $\pi$-stacking.

(d) In situations with significant nondynamical correlation (particularly type A), where a single HF determinant is a poor zero-order representation of the wave function, HF-DFT inherits this problem and is actually detrimental. In this context, the work on multiconfiguration pair DFT,[86,87] which predates recent interest in HF-DFT, should be mentioned for perspective.

(e) If one is *determined* to use nonempirical functionals, HF-SCAN-D4 appears to be worth considering, with a WTMAD2 for a *nonempirical* functional that compares favorably with hybrid GGAs and meta-GGAs except for the heavily parameterized M06-2X[88,89] (WTMAD2 = 4.79 kcal/mol) and a cost that is marginally greater than a simple HF calculation.

# ■ ASSOCIATED CONTENT

**ⓈSupporting Information**

The Supporting Information is available free of charge at https://pubs.acs.org/doi/10.1021/acs.jctc.0c01055.

Abridged details of all 55 subsets of GMTKN55 with proper references, optimized D4 parameters, and five major subcategory contributions for total WTMAD2 for both HF and self-consistent PBE$n$, B$n$LYP, SCAN$n$, and TPSS$n$ functionals, with and without dispersion; dependence of subset-wise WTMAD2 contributions on the percentage of HF exchange for the SCAN$n$-D4 and B$n$LYP-D4 series ($n$ ranging from 0 to 50%); dependence of WTMAD2 (kcal/mol) and the five top-level subsets on the percentage of HF exchange ($x$-axis) for dispersion-free HF as well as self-consistent PBE$n$, B$n$LYP, SCAN$n$, and TPSS$n$ series; effect of considering D4 dispersion correction on WTMAD2 (kcal/mol) and the five top-level subsets for HF-SCAN$n$ and SCAN$n$ series; and sample Gaussian 16 and ORCA 4 inputs for HF-DFT (PDF)

# ■ AUTHOR INFORMATION

**Corresponding Author**

Jan M.L. Martin − *Department of Molecular Chemistry and Materials Science, Weizmann Institute of Science, 7610001 Reḥovot, Israel;* orcid.org/0000-0002-0005-5074; Email: gershom@weizmann.ac.il

**Author**

Golokesh Santra − *Department of Molecular Chemistry and Materials Science, Weizmann Institute of Science, 7610001 Reḥovot, Israel;* orcid.org/0000-0002-7297-8767

Complete contact information is available at:
https://pubs.acs.org/10.1021/acs.jctc.0c01055

**Funding**

This research was supported by the Israel Science Foundation (grant 1969/20) and by the Minerva Foundation, Munich, Germany (grant 20/05). GS acknowledges a fellowship from the Feinberg Graduate School of the Weizmann Institute.

**Notes**

The authors declare no competing financial interest.

# ■ ACKNOWLEDGMENTS

The authors would like to thank Drs. Irena Efremenko and Mark A. Iron (both at Weizmann) for helpful discussions, Dr. Christoph Bannwarth (Stanford U.) for feedback on a draft manuscript, and Dr. Mark Vilensky (scientific computing manager of CHEMFARM) for technical assistance.